\documentclass[%
reprint,
amsmath,amssymb,
aps,
pra,
]{revtex4-1}

\usepackage{graphicx,wrapfig,lipsum}
\usepackage{subeqnarray}
\usepackage{url,hyperref} 
\usepackage{color}
\usepackage{ulem}
\usepackage{verbatim}

\usepackage{natbib}

\newcommand{\req}[1]{Eq.\,({\ref{#1}})}

\begin{document}
%
\title{Motion of classical charged particles with magnetic moment in external plane-wave electromagnetic fields}
\author{Martin Formanek}
\email{martinformanek@email.arizona.edu}
\author{Andrew Steinmetz}
\email{ajsteinmetz@arizona.edu}
\author{Johann Rafelski}
\email{johannr@arizona.edu}
\affiliation{Department of Physics,
The University of Arizona,
Tucson, AZ, 85721, USA}

\date{\today}
%
\begin{abstract}
We study the motion of a charged particle with magnetic moment in external electromagnetic fields utilizing covariant unification of Gilbertian and Amperian descriptions of particle magnetic dipole moment. Considering the case of a current loop, our approach is verified by comparing classical dynamics with the classical limit of relativistic quantum dynamics. 
We obtain motion of a charged particle in the presence of an external linearly polarized EM (laser) plane wave field incorporating the effect of spin dynamics. For specific laser-particle initial configurations, we determine that the Stern-Gerlach force can have a cumulative effect on the trajectory of charged particles. 
\begin{description}
\item[PACS numbers]
13.40.Em Electric and magnetic moments, 
03.30.+p Special relativity 
\end{description}
\end{abstract}
\maketitle


\section{Introduction}
\noindent In the context of high intensity laser-matter interaction, including particle acceleration, much attention is being paid to classical dynamics of charged particles, in particular electrons and positrons. We consider here the contribution of the Stern-Gelrach force due to magnetic moment, further advancing the work of Wen, Keitel and Bauke~\cite{Wen:2017zer}. Since this force is much smaller compared to Lorentz force, and a well defined dynamical formulation was presented only recently~\cite{Rafelski:2017hce}, much work remains to be done. 

Our theoretical formulation is building upon this covariant unification of Amperian and Gilbertian dynamics and the study of {\it neutral} particle dynamics in presence of the Stern-Gerlach force~\cite{Formanek:2018mbv,Formanek:2019cga}. Study of Stern-Gerlach particle dynamics along this conceptual approach was initiated by Good~\cite{Good}, and Nyborg~\cite{Nyborg},  see Ref.~\cite{Bagrov1980} for review of this work.

We begin by presenting the formulation of the model for particle motion and spin dynamics in Section~\ref{sec:dynamics}, showing how the magnetic moment force effect on the particle's translational motion is included. Our Stern-Gerlach force is a natural extension of the Thomas-Bargmann-Michele-Telegdi (TBMT) spin precession dynamics~\cite{Thomas1927,Bargmann:1959gz}. It shows how the magnetic moment interacts with the external electromagnetic (EM) field inhomogenities and how this interaction affects the particle's trajectory. 

In Section~\ref{sec:currentloop}, we check our approach validity by comparing with Ref.\,\cite{Wen:2017zer} for the case of a charged particle with a magnetic moment moving across a current loop. In Section~\ref{sec:planewave}, we turn to our main objective, the study of dynamics in presence of (laser) plane waves. To solve this intricate dynamical problem, we adopt covariant techniques employed in the study of exact charged spin-0 particle dynamics~\cite{Sarachik1970} and further developed in the study of radiation reaction effects~\cite{Dipiazza2008,Hadad:2010mt}. 

These methods allow us to identify and use conservation laws along with differential equations for the covariant projections to reduce the coupled equations for particle motion and spin dynamics to a greatly simplified and analytically solvable set. Our solution including spin dynmics is analytical and transparent, allowing applications to environments where magnetic moment dynamics could be relevant.

We summarize, discuss, and evaluate the achieved results in Section~\ref{Sec:Final},

\section{Dynamics of Charged particle with magnetic moment}\label{sec:dynamics}

The covariant and unified (Amperian=Gilbertian) \lq dipole charge model\rq\ was formulated by us in Ref.~\cite{Rafelski:2017hce} and previously in~\cite{Good,Nyborg}. The magnetic moment interaction is incorporated through the \lq magnetic 4-potential\rq\ $B^\mu$
\begin{equation}\label{eq:4potential}
B_\mu \equiv F^*_{\mu\nu}s^\nu, \quad \text{where} \quad F^*_{\mu\nu} \equiv \frac{1}{2}\varepsilon_{\mu\nu\alpha\beta}F^{\alpha\beta}\,,
\end{equation}
with $\varepsilon_{0123} = +1$ is a dual tensor to the EM tensor $F^{\mu\nu}$. The 4-potential $B_\mu$ was constructed~\cite{Rafelski:2017hce} in such a way that in the co-moving frame $u^\mu = (c,0)$ a following quantity
\begin{equation}
d_m B \cdot u = cd_m F^*_{0\nu}s^\nu = -\pmb{\mu}_m \cdot \pmb{\mathcal{B}}\,.
\end{equation}
gives us the correct potential energy of an elementary magnetic moment $\pmb{\mu}_m$ in an external magnetic field $\pmb{\mathcal{B}}$. We have introduced the conserved \lq magnetic dipole charge\rq\, $d_m$: in the rest frame of the particle it has the meaning of a proportionality constant between the magnetic moment and particle spin
\begin{equation}\label{eq:magmoment}
\pmb{\mu}_m = c d_m \pmb{s}, \quad cd_m = \frac{e}{m} + \tilde{a}\;,
\end{equation}
where $\tilde{a} = a e /m$ is proportional to anomalous magnetic moment $a$. For electrons $a \approx \alpha / 2 \pi = 1.16\times 10^{-3}$. 

With the 4-potential $B^\mu$, we can formulate the equation of motion as
\begin{equation}
m \dot{u}^\mu = e F^{\mu\nu}u_\nu + d_m G^{\mu\nu}u_\nu\,,
\end{equation}
where the \lq dot\rq\ denotes a derivative with respect to proper time $\tau$. The tensor $G_{\mu\nu}$ reads
\begin{equation}
G_{\mu\nu} \equiv \partial_\mu B_\nu - \partial_\nu B_\mu\,.
\end{equation}
Substituting the definition of magnetic 4-potential $B_\mu$ from Eq.\,\eqref{eq:4potential} and performing usual algebra to obtain regular Lorentz force terms, we arrive to
\begin{multline}
\dot{u}^\mu = \frac{e}{m}F^{\mu\nu}u_\nu - \frac{d_m}{m} u \cdot \partial(F^{*\mu\nu}s_\nu)\\
 + \frac{d_m}{m}\partial^\mu(u \cdot F^* \cdot s)\,.
\end{multline}
The last two terms can be simplified using the the covariant Maxwell equation
\begin{equation}
\partial_\mu F^*_{\alpha\beta} + \partial_\alpha F^*_{\beta\mu} + \partial_\beta F^*_{\mu\alpha} = 0\,,
\end{equation}
while considering that the partial derivatives commute with the $u^\mu$ and $s^\mu$ 4-vectors, since the derivatives act only on the field quantities. We are left with the equation of motion
\begin{equation}\label{eq:udynam2}
\dot{u}^\mu = \frac{1}{m} \left(e F^{\mu \nu} 
- d_m s \cdot \partial F^{*\mu\nu}\right)u_\nu\,.
\end{equation}
The first term on the RHS of Eq.\,\eqref{eq:udynam2} is the standard Lorentz force, while the second is the covariant version of the Stern-Gerlach force term. Since the magnetic 4-potential $B^\mu$ is gauge invariant, a $d_m B \cdot u$ term can be added with impunity to the $eA\cdot u$ term in the Lagrangian action, resulting in the variational principle origin of Eq.\,\eqref{eq:udynam2}, up to sub-leading higher order spin dynamics terms arising from $ds^\mu/d\tau$ term in Euler-Lagrange equations.

Now using Schwinger's method~\cite{schwinger1974}, we construct the spin dynamics equations applying the constraints
\begin{align} 
u \cdot s = 0, \quad &\Rightarrow \quad \dot{u} \cdot s + u \cdot \dot{s} = 0\,,\label{eq:us}\\ 
s^2 = \text{const}, \quad &\Rightarrow \quad s \cdot \dot{s} = 0\label{eq:s2}\,.
\end{align} 
The exact solution linear in external fields for the dynamics considered in Eq.\,\eqref{eq:udynam2} is
 \begin{multline}\label{eq:spindynam}
\dot{s}^\mu = \frac{e}{m}F^{\mu\nu}s_\nu + \widetilde{a} \left( F^{\mu\nu}s_\nu - \frac{u^\mu}{c^2} u \cdot F \cdot s\right)\\ 
- \frac{d_m}{m} s \cdot \partial F^{*\mu\nu}s_\nu\;.
\end{multline}
The first term assures consistency with the Lorentz force component, Eq.\,\eqref{eq:us}, the second term encompasses anomalous magnetic moment behavior, and the third ensures consistency with the Stern-Gerlach force term in Eq.\,\eqref{eq:udynam2}. Equation~\eqref{eq:spindynam} is a \lq minimal\rq\ solution of the Schwinger consistency requirements, in a sense that additional terms preserving the conditions Eq.\,\eqref{eq:us} and Eq.\,\eqref{eq:s2} are possible, but not necessary without additional physical requirements. 
\section{Particle in an inhomogeneous magnetic field}\label{sec:currentloop}
\noindent We consider a magnetic field pointing along the $z$-axis $\pmb{\mathcal{B}} = \mathcal{B}_z(z) \hat{z}$. 
The initial 4-velocity and 4-spin are oriented along the $z$-axis as well
\begin{align}
u^\mu(0) &= \gamma_0 c(1, 0,0,\beta_0)\,,\\
s^\mu(0) &= \gamma_0 s_0 (\beta_0, 0,0,1)\,,	
\end{align}
where $s_0 = \pm \hbar/2$ is positive for spin oriented along the positive $z$-axis or negative for the opposite case. Initially there is no Lorentz force on the particle since it moves parallel to the magnetic field. In fact, in this configuration the motion remains 1D because all products $F^{\mu\nu}u_\nu$ and $F^{\mu\nu}s_\nu$ start at zero and remain zero because the Stern-Gerlach terms contribute only to zeroth and $z$-components. We can then effectively rewrite the equations~\eqref{eq:udynam2} and \eqref{eq:spindynam} as
\begin{align}
\label{eq:dynamicloop}\dot{u}^\mu &= - \frac{d_m}{m} s \cdot \partial (F^{*\mu\nu})u_\nu\,,\\
\label{eq:torqueloop}\dot{s}^\mu &= - \frac{d_m}{m} s \cdot \partial (F^{*\mu\nu})s_\nu\,,
\end{align}
The torque Eq.\,(\ref{eq:torqueloop}) in this case does not change the direction of the spin, only its velocity dependence is modified so that $u\cdot s = 0$ is satisfied. From this argument alone the solution for spin is
\begin{equation}
s^\mu(\tau) = \gamma s_0 (\beta, 0,0,1)\,,
\end{equation}
where $\gamma$ is the relativistic Lorentz factor. The consistent solution for the change of particle velocity $\beta$ as a function of position can be derived from either eqs.~(\ref{eq:dynamicloop},\ref{eq:torqueloop}) as
\begin{equation}
\frac{d\beta}{dt} = \frac{d_ms_0}{m}\frac{\partial_z \mathcal{B}_z(z)}{\gamma^2}\,.
\end{equation}
Without magnetic moment $d_m = 0$ the particle would pass through the region of the magnetic field unimpeded with constant velocity equal to its initial velocity $\beta_0$. The Stern-Gerlach force on magnetic moment accelerates or decelerates the particle based on the direction of the spin $s_0$ and sign of the gradient $\partial_z B_z$. 

\begin{figure}
	\includegraphics[width=\linewidth]{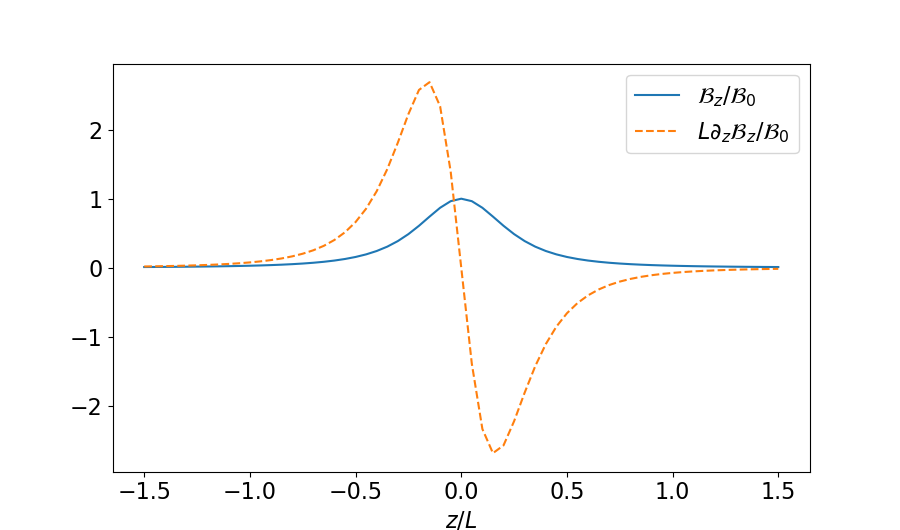}
	\caption{\label{fig:bfield}Magnetic field and its derivative along the axis of a current loop with radius $L/\pi$. See Eq.\,\eqref{eq:bfield}.}
\end{figure}

\begin{figure}
	\includegraphics[width=\linewidth]{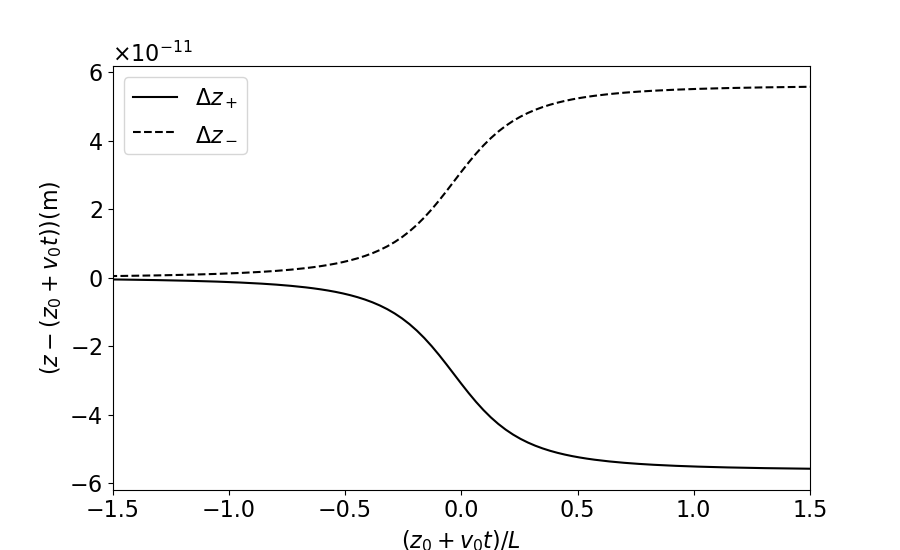}
	\caption{\label{fig:trajectory}Numerical solutions of both aligned spin up $(+)$ and anti-aligned spin down $(-)$ electron motion in the magnetic field from Figure~\ref{fig:bfield} with $\mathcal{B}_0 = 10\ \text{T}$ and $L/\pi = 1\ \text{cm}$. The spin is measured with respect to positive $z$-axis. Plotted are the differences from the trajectory of a reference electron moving with constant velocity $v_0 = 2 \times 10^8\ \text{m/s}$ with no magnetic field.}
\end{figure}

As in Ref.~\cite{Wen:2017zer} we will model the magnetic field as the field generated by a current loop with a radius $L/\pi$. Along the $z$-axis, the magnetic field has the form 
\begin{equation}\label{eq:bfield}
\mathcal{B}_z = \frac{\mathcal{B}_0}{(1 + \pi^2 z^2 / L^2)^{3/2}}\,.
\end{equation}
The radial component of the magnetic field $\mathcal{B}_r$ vanishes on the axis although its derivative $\partial_r$ is non-zero, thus satisfying Maxwell equation $\nabla \cdot \pmb{\mathcal{B}} = 0$  (See p. 181 in Ref.~\cite{jackson}). The derivative $\partial_z B_r$ is zero on the axis, ensuring that for perfectly polarized electrons the motion remains 1D. The field and its derivative are plotted in Figure~\ref{fig:bfield}. For electrons $d_m < 0$ since the direction of their magnetic moment is opposite to spin. Thus electrons entering the field from the left with aligned spin $s_0 = + \hbar/2$ would first be slowed down by the increasing field and then accelerated back again as the particle leaves. This is consistent with the textbook Stern-Gerlach force $\pmb{F} = \nabla(\pmb{\mu}_m \cdot \pmb{\mathcal{B}})$. The velocity of particle with aligned spin is thus smaller than $\beta_0$ throughout the motion. The anti-aligned spin $s_0 = - \hbar/2$ electron would first be accelerated and then decelerated so that its velocity is greater than $\beta_0$ throughout the motion. This means that electrons with aligned spin (+) would lag behind electrons with anti-aligned spin ($-$) when moving through the same region. We can compare their trajectories with the motion of electrons when the magnetic field is absent using
\begin{equation}
\Delta z_\pm = z_\pm(t) - (z_0 + v_0 t)\,.
\end{equation}
The plots for the numerical solutions with maximum magnetic field strength $\mathcal{B}_0 = 10\ \text{T}$ and radius $L/\pi = 1\ \text{cm}$ are presented in Figure~\ref{fig:trajectory}. The numerical integration was initialized at time $t_0 = 0$ for the electron position $z_0/L = - 5.0$ and initial velocity $v_0 = 2 \times 10^8\ \text{m/s}$. Qualitatively, the spread in distance between the two oppositely polarized electron types
\begin{equation}
\Delta z \equiv \Delta z_- - \Delta z_+ \sim \frac{1}{\gamma^2}\,,
\end{equation}
has the same behavior found from the Foldy-Wouthuysen model tracking the classical limit of quantum magnetic moment dynamics. See discussion in Section 3.2 in~\cite{Wen:2017zer} which shows the same $1/\gamma^2$ dependence unlike the classical model they used with $1/\gamma$ behavior. As was also pointed out in~\cite{Wen:2017zer}, distinguishing between magnetic moment models based on this experiment is a challenge. This is because the difference becomes substantial only for high gamma factors when the flight time of electrons is much shorter and thus the trajectory differences due to the magnetic moment interaction decrease. Any experiment would be also limited by how well the electrons can be polarized along the $z$-axis and by their displacement from this axis.

\section{Charged particle in linearly polarized plane wave field}\label{sec:planewave}
\subsection{Problem definition}
\noindent In our previous work~\cite{Formanek:2019cga}, we presented an analytical solution for neutral particle motion in an external plane wave field based on the covariant approach of~\cite{Dipiazza2008,Hadad:2010mt}. Here, the situation is more complicated because the particle is charged and feels a corresponding Lorentz force. Nevertheless, we will demonstrate that an analytical solution can still be found. 

In the neutral particle case there is no Lorentz force, and the magnetic moment interaction is a first order effect. Previously, we discussed such interactions for hypothetical neutrino magnetic moments and neutron beam control~\cite{Formanek:2018mbv}, which required enormous field intensities to produce a measurable effect. Electrons have an advantage of having orders of magnitude higher magnetic moment than all other stable charged particles, neutrons, or the yet-to-be-found magnetic moment of neutrinos. Thus, we look here specifically at electron dynamics allowing the Stern-Gerlach force to affect their trajectory to a much greater extent than is the case for all other particles.

The relativistic effects for particles in external laser fields are controlled by a Lorentz-invariant parameter $a_0$, so-called unitless normalized laser amplitude~\cite{Mourou2006}, given by
\begin{equation}\label{eq:a0}
a_0 = \frac{e\sqrt{|A^\mu A_\mu|}}{mc} = \frac{e\mathcal{E}\lambdabar}{mc^2} = 1.2 \frac{\sqrt{I[10^{20} \text{ W/cm}^2]}}{\hbar \omega[1 \text{ eV}]}\,,
\end{equation}
where $I$ is intensity of the laser and $\omega$ its frequency. This quantity corresponds to the work done by a laser's electric field $\mathcal{E}$ over one reduced wavelength $\lambdabar$ compared to the particle's rest mass energy $mc^2$. For $a_0 \sim 1$ we enter the relativistic regime and for $a_0 \gg 1$ the ultra-relativistic regime.

\subsection{Classical vs quantum dynamics}\label{sec:clasvsquant}

The classical particle dynamics should arise as a limit of relativistic quantum theory. The most common approach to deriving these equations in the classical limit relies on a Foldy-Wouthuysen transformation~\cite{Foldy1950} of the spin-1/2 Dirac equation. This is followed by introducing a correspondence principle for the time evolution of observables such as position, kinematic momentum, and spin in the Heisenberg picture~\cite{Silenko:2007wi}. 

In the situation of strong external EM fields for particles with gyromagnetic ratio $g\ne 2$, we have argued the correct quantum relativistic description of particle dynamics is not necessarily based on the Dirac equation. As we pointed out in Ref.\,\cite{Steinmetz:2018ryf}, a more natural approach to anomalous magnetic moment is found in the Klein-Gordon-Pauli (KGP) equation which incorporates a Pauli term, capturing the dynamics of magnetic moment, into Klein-Gordon equation. 

This insight is not compatible with currently most common approach, the use the Dirac-Pauli (DP) equation where the Dirac equation is supplemented by a Pauli term. The primary difference between these two approaches is that while in KGP the entire magnetic moment, and thus spin dynamics, is described by a single mathematical object, the DP approach breaks apart the magnetic moment into a natural $g=2$ part embedded in the spinor structure and an anomalous part described by the Pauli term. 

In prior work~\cite{Steinmetz:2018ryf} we presented an argument that the difference between these two approaches becomes apparent in strong EM fields which can be found around magnetars or in high-$Z$ atoms. However, for the external fields considered in this work the difference between DP and KGP formulation will not be apparent in consistency with our prior assumption to neglect subleading spin dynamics effects, see Section~\ref{sec:dynamics}.

The parameter space controlling the classical domain is described in Ref.~\cite{Khokonov}. Apart from the normalized laser amplitude $a_0$ \req{eq:a0} we invoke a Lorentz invariant parameter 
\begin{equation}\label{eq:aq}
a_q = \frac{\hbar (k \cdot p) }{m^2c^2}\,,
\end{equation}
where $k^\mu$ and $p^\mu$ are 4-momenta of the photon and electron respectively. In this section we consider the example of the electron at rest $a_q = \hbar \omega / mc^2 \approx 10^{-6}$ for 1eV visible laser light. From the diagram in Ref.~\cite{Khokonov} we see that any $a_0$ satisfying $\ln a_0 < 4$ allows us to treat the problem classically. Later we consider example of $a_0 = 0.1$ which is squarely in the classical domain. In this work we don't consider the radiation of the electrons due to their motion.

In the literature one often sees the Lorentz invariant parameter $\chi$ as defined by~\cite{Ritus1985} 
\begin{multline}\label{eq:chi}
\chi = \frac{e\hbar \sqrt{|u\cdot F \cdot F \cdot u|}}{m^2c^3} = \left. \frac{\mathcal{E}}{\mathcal{E}_\text{S}}\right|_{CF} = a_0 a_q \\
=  5.9\times 10^{-2} E[\text{GeV}] \sqrt{I[10^{20} \text{W/cm}^2]}\,,
\end{multline}
where $E$ is the electron energy. This parameter represents the electric field strength $\mathcal{E}$ in units of the critical Schwinger field in the co-moving frame of the electron. For an electron $\mathcal{E}_S = 1.3\times 10^{18}$ V/m. This parameter is a product of the two previous ones \req{eq:a0} and \req{eq:aq}. It has been shown that quantum effects become non-negligible already for $\chi \gtrsim 0.1$~\cite{Uggerhoj}.

\subsection{Dynamical equations}
The covariant 4-potential for a plane wave field is given by
\begin{equation}\label{eq:plane wave}
A^\mu = \varepsilon^\mu \mathcal{A}_0 f(\xi), \quad \xi = \frac{\omega}{c}\hat{k} \cdot x\;,
\end{equation}
where the unitless wave 4-vector $\hat{k}^\mu$ is light-like and orthogonal to the space-like polarization vector $\varepsilon^\mu$. We impose the following constraints which are satisfied by plane waves
\begin{equation}\label{eq:properties}
\quad \hat{k}^2 = 0, \quad \hat{k} \cdot \varepsilon = 0, \quad \varepsilon^2 = -1\,. 
\end{equation}
The amplitude of the field is given by $\mathcal{A}_0$ and $\xi$ denotes its invariant phase. The oscillatory part of the laser field and the pulse envelope are then defined by some function given by $f(\xi)$ unique to the laser.

Substituting the 4-potential into the EM field tensor yields
\begin{equation}\label{eq:emtensor}
F^{\mu\nu} = \partial^\mu A^\nu - \partial^\nu A^\mu = \frac{\mathcal{A}_0 \omega}{c}f'(\xi) (\hat{k}^\mu \varepsilon^\nu - \varepsilon^\mu \hat{k}^\nu)\;.
\end{equation}
In our notation, prime marks (such as $f'$) are used to denote derivatives with respect to the phase $\xi$. 

The properties of Eq.\,\eqref{eq:properties} ensure that the contraction of the $F^{\mu\nu}$ tensor with $\hat{k}^\mu$ is zero. It is also useful to calculate
\begin{equation}\label{eq:dualtensor}
(s\cdot \partial)F^{*\mu\nu} = \frac{\mathcal{A}_0\omega^2}{c^2} f''(\xi) (\hat{k} \cdot s) \epsilon^{\mu\nu\alpha\beta}\hat{k}_\alpha\varepsilon_\beta \;,
\end{equation}
since this term appears in both particle and spin dynamics equations \eqref{eq:udynam2} and \eqref{eq:spindynam}. Notice that contracting Eq.\,\eqref{eq:dualtensor} with both $\hat{k}_\mu$ and $\varepsilon_\mu$ yields zero because of the antisymmetric properties of $\epsilon^{\mu\nu\alpha\beta}$. This means that in the projections with $\hat{k}^\mu$ and $\varepsilon^\mu$ the Stern-Gerlach term does not play a role. 

The dynamical equations \eqref{eq:udynam2} and \eqref{eq:spindynam} in terms of the plane wave potential (\ref{eq:plane wave}) are then
\begin{align}
\label{eq:firstplanewave}
\dot{u}^\mu &= \frac{e\mathcal{A}_0 \omega}{mc}f'(\xi)(\hat{k}^\mu (\varepsilon \cdot u) - \varepsilon^\mu (\hat{k} \cdot u)) \nonumber\\
&- \frac{\mathcal{A}_0d_m\omega^2}{mc^2} f''(\xi) (\hat{k} \cdot s) \epsilon^{\mu\nu\alpha\beta}u_\nu \hat{k}_\alpha\varepsilon_\beta\;,\\
\dot{s}^\mu &= \omega d_m \mathcal{A}_0 f'(\xi) (\hat{k}^\mu \varepsilon \cdot s - \varepsilon^\mu \hat{k} \cdot s) - u^\mu (u \cdot F \cdot s) \frac{\widetilde{a}}{c^2} \nonumber \\
&-\frac{\mathcal{A}_0 d_m \omega^2}{mc^2} f''(\xi) (\hat{k} \cdot s) \epsilon^{\mu\nu\alpha\beta}s_\nu \hat{k}_\alpha\varepsilon_\beta\label{eq:secondplanewave}\;.
\end{align}
These two equations are coupled through the Stern-Gerlach interaction. Namely the quantity of interest is the function $\hat{k}\cdot s(\tau)$ which appears in the coupling term. In the following Section (\ref{sec:projections}) we present an analytical solution for this function and in Section (\ref{sec:4velocity}) we find the analytical solution for 4-velocity $u^\mu(\tau)$.

\subsection{Spin precession in $\hat{k}^\mu$ and $\varepsilon^\mu$ projections}\label{sec:projections}
\noindent As we mentioned above, the coupling through the Stern-Gerlach force between the equation of motion Eq.\,\eqref{eq:firstplanewave} and spin dynamics Eq.\,\eqref{eq:secondplanewave} disappears if projected with $\hat{k}^\mu$ and $\varepsilon^\mu$. In such a case the motion and spin precession is governed by only the TBMT equations of motion~\cite{Bargmann:1959gz}. The situation here is more complex than the neutral particle case presented in~\cite{Formanek:2019cga} as the equations of motion gain additional terms proportional to particle charge. In the charged case, the projection of 4-velocity and laser polarization $\varepsilon \cdot u(\tau)$ is no longer a constant of motion. Instead we obtain after multiplying Eq.\,\eqref{eq:firstplanewave} by $\varepsilon^\mu$
\begin{equation}
\varepsilon \cdot \dot{u} = \frac{d}{d\tau} (\varepsilon \cdot u) = \frac{e\mathcal{A}_0 \omega}{mc} k \cdot u(0) f'(\xi)\;,
\end{equation}
which can be integrated as
\begin{equation}\label{eq:secondintegral}
\varepsilon \cdot u(\tau) = \varepsilon \cdot u(0) + \frac{e}{m} A_0 (f(\xi(\tau)) - f(\xi_0))\;.
\end{equation}
The projection $\epsilon\cdot u(\tau)$ becomes sensitive to the laser profile as a function of the laser phase and is proportional to $e/m$.

Similarly to the neutral particle case, the charged particle's projection of wave 4-vector and initial 4-velocity $\hat{k}\cdot u(0)$ remains a constant of motion which can be seen by multiplying Eq.\,\eqref{eq:firstplanewave} by $\hat{k}^\mu$ yielding 
\begin{equation}\label{eq:firstintegral}
\hat{k} \cdot \dot{u} = \frac{d}{d\tau} (\hat{k} \cdot u) = 0, \quad \Rightarrow \quad \hat{k} \cdot u = \hat{k} \cdot u(0)\;.
\end{equation}
This expression also allows us to find the relationship between the proper time of the particle and phase of the wave $\xi$ as 
\begin{equation}\label{eq:phase}
\frac{d\xi}{d\tau} = \frac{\omega}{c}\frac{d}{d\tau}(\hat{k} \cdot x) = \frac{\omega}{c}\hat{k} \cdot u(0), \; \Rightarrow \; \xi = \frac{\omega}{c}(\hat{k} \cdot u(0)) \tau + \xi_0\;.
\end{equation}

Now we turn our attention to the spin dynamics of Eq.\,\eqref{eq:secondplanewave}. Using the integral of motion Eq.\,\eqref{eq:firstintegral} we can evaluate the contraction of torque with $\hat{k}^\mu$ resulting in
\begin{equation}\label{eq:kprod}
\hat{k} \cdot \dot{s} = - (\hat{k} \cdot u(0)) (u \cdot F \cdot s)\frac{\widetilde{a}}{c^2}\,.
\end{equation}
By taking another proper time derivative of this equation and after some algebra which also uses the projection $\varepsilon \cdot \dot{s}$ (for details see~\cite{Formanek:2019cga}) we arrive at second order differential equation for the function $\hat{k}\cdot s(\tau)$
\begin{equation}
\label{Eq:SpinProjection}
\hat{k} \cdot \ddot{s} = \frac{\ddot{f}(\xi(\tau))}{\dot{f}(\xi(\tau))} (\hat{k} \cdot \dot{s}) - \frac{\widetilde{a}^2\mathcal{A}_0^2}{c^2} \dot{f}^2(\xi(\tau))(\hat{k} \cdot s)\;.
\end{equation}
We introduce a set of known initial conditions 
\begin{align}
\hat{k} \cdot s(\tau = 0) &= \hat{k} \cdot s(0)\,,\\
\hat{k} \cdot \dot{s}(\tau = 0) &= - (\hat{k}\cdot u(0)) (u(0)\cdot F \cdot s(0))\frac{\widetilde{a}}{c^2}\;.
\end{align}
which can be used to construct a solution
\begin{equation}\label{eq:kssolution}
\hat{k} \cdot s(\tau) = \hat{k} \cdot s(0) \cos \left[\frac{\widetilde{a}\mathcal{A}_0}{c}\psi(\tau)\right] - \frac{W(0)}{c} \sin \left[\frac{\widetilde{a}\mathcal{A}_0}{c}\psi(\tau)\right]\;,
\end{equation}
in which we for simplicity write the difference between the laser amplitude at a given time from the initial as
\begin{equation}\label{eq:psi}
\psi(\tau) \equiv f(\xi(\tau)) - f(\xi_0)\,.
\end{equation}
We note that $\psi(\tau)$ is also present in Eq.\,\eqref{eq:secondintegral}. The function $W(\tau)$ is proportional to $u \cdot F \cdot s$ and is defined as
\begin{equation}\label{eq:wdef}
W(\tau) \equiv (\hat{k} \cdot u(0)) (\varepsilon \cdot s(\tau)) - (\varepsilon \cdot u(\tau)) (\hat{k} \cdot s(\tau))\;.
\end{equation}
Analogously, we can obtain the solution for $\varepsilon \cdot s(\tau)$ as 
\begin{multline}\label{eq:epsilondotssolution}
\varepsilon \cdot s(\tau) = \left(\varepsilon \cdot s(0) + \frac{\hat{k}\cdot s(0)}{\hat{k} \cdot u(0)}\frac{e\mathcal{A}_0}{m}\psi(\tau) \right)\cos \left[\frac{\widetilde{a}\mathcal{A}_0}{c}\psi(\tau)\right]\\
+ \left(\frac{c \hat{k} \cdot s(0)}{\hat{k} \cdot u(0)} - \frac{W(0)}{c} \frac{\varepsilon \cdot u(\tau)}{\hat{k} \cdot u(0)} \right) \sin \left[\frac{\widetilde{a}\mathcal{A}_0}{c}\psi(\tau)\right]\;.
\end{multline}
Note that the spin projection precession is governed only by the value of the anomalous magnetic moment $\widetilde{a} = a e /m$. For no magnetic anomaly the precession vanishes
\begin{align}\label{eq:ksnomag}
\hat{k}\cdot s(\tau) &= \hat{k}\cdot s(0)\,,\\
\varepsilon \cdot s(\tau) &= \varepsilon \cdot s(0) + \frac{\hat{k}\cdot s(0)}{\hat{k}\cdot u(0)} \frac{e\mathcal{A}_0}{m}\psi(\tau)\label{eq:epssnomag}\,.
\end{align} 
Let us finally consider a particle with an initial configuration ($\tau = 0$) long before the arrival of the pulse such that the envelope function is $f(\xi_0) = 0$. Then long after the pulse leaves $\psi(\tau\rightarrow \infty) = 0$ the projections of wave 4-vector and polarization on spin should relax back to their original values
\begin{align}
\hat{k} \cdot s(\tau \rightarrow \infty) &= \hat{k} \cdot s(0)\,,\\
\varepsilon \cdot s(\tau \rightarrow \infty) &= \varepsilon \cdot s(0)\,.
\end{align}
These parameters are only reversibly altered by the presence of a plane wave, excluding any deviations that would arise if the particle radiates due its motion, effect not considered in this work.

\subsection{Particle 4-velocity $u^\mu(\tau)$}\label{sec:4velocity}
Our ultimate goal in discussing this test case of the charged particle with spin under the influence of a plane wave is to derive how the particle's trajectory and motion are altered, especially by the presence of the anomalous magnetic moment. Our first step in deriving the 4-velocity directly is to construct another integral of motion by considering the 4-vector
\begin{equation}\label{eq:lmu}
L^\mu \equiv \epsilon^{\mu\nu\alpha\beta}u_\nu(0)\hat{k}_\alpha \varepsilon_\beta
\end{equation} 
and projecting the equation of motion Eq.\,\eqref{eq:firstplanewave} along the direction of $L^{\mu}$ yielding
\begin{equation}\label{eq:lutau}
L \cdot \dot{u}(\tau) = \frac{d_m \mathcal{A}_0 \omega^2}{mc^2} f''(\xi) (\hat{k}\cdot s) (\hat{k}\cdot u(0))^2\,,
\end{equation}
where we used the contraction identity 
\begin{equation}\label{eq:contraction}
\epsilon^{\mu\nu\alpha\beta} \epsilon_{\mu\rho\gamma\delta} = - \left| \begin{matrix}
\delta^\nu_\rho & \delta^\nu_\gamma & \delta^\nu_\delta \\
\delta^\alpha_\rho & \delta^\alpha_\gamma & \delta^\alpha_\delta \\
\delta^\beta_\rho & \delta^\beta_\gamma & \delta^\beta_\delta\\
\end{matrix} \right| \,,
\end{equation}
and the constant of motion Eq.\,\eqref{eq:firstintegral}. Equation~\eqref{eq:lutau} can be formally integrated with initial condition $L\cdot u(0) = 0$ which is true due to the antisymmetry of $\epsilon^{\mu\nu\alpha\beta}$ in Eq.\,\eqref{eq:lmu}. This results in
\begin{equation}\label{eq:Lu}
L \cdot u(\tau) = - h(\tau) (\hat{k}\cdot u(0))^2\,.
\end{equation}
The unitless integral $h(\tau)$ is given by
\begin{equation}\label{eq:htau}
h(\tau) \equiv - \frac{d_m\mathcal{A}_0\omega^2}{mc^2}\int_{\tau_0 = 0}^\tau \hat{k} \cdot s(\widetilde{\tau}) f''(\xi(\widetilde{\tau}))d\widetilde{\tau}\,,
\end{equation}
and depends on the known solution for the spin projection $\hat{k}\cdot s(\tau)$ from Eq.\,\eqref{eq:kssolution}. For a constant $\hat{k}\cdot s(\tau) = \hat{k}\cdot s(0)$ which is realized in the case of no magnetic anomaly $\widetilde{a} = 0$ this integral can be evaluated as 
\begin{equation}
h(\tau) = - \frac{d_m \mathcal{A}_0 \omega}{mc} \frac{\hat{k}\cdot s(0)}{\hat{k}\cdot u(0)}[f'(\xi(\tau)) - f'(\xi_0)]\,,
\end{equation}
satisfying the initial condition $h(\tau = 0) = 0$. In this situation the function $h(\tau)$ is oscillatory as it is proportional to the derivative $f'(\xi(\tau))$. Thus, the value of $h(\tau)$ for no magnetic anomaly doesn't accumulate over the interaction with many plane wave cycles. 

This function is responsible for irreversible effects in which the particle is changed after the passage of an EM plane wave, but only in the situation that an anomalous magnetic moment is present. In the next section (Section~\ref{sec:restparticle}) we will discuss under which circumstances the integral for $h(\tau)$ (\ref{eq:htau}) is cummulative for particle initially at rest.

We will look for the 4-velocity by assuming an ansatz 
\begin{equation}\label{eq:ansatz}
u^\mu(\tau) = u^\mu(0) + C_1(\tau)\varepsilon^\mu + C_2(\tau)\hat{k}^\mu + C_3(\tau) L^\mu\,.
\end{equation}
The norm of the last 4-vector is manifestly negative
\begin{equation}
L^2 = - (\hat{k}\cdot u(0))^2\,,
\end{equation}
and therefore this 4-vector is space-like.

The solution ansatz Eq.\,\eqref{eq:ansatz} automatically preserves the projection $\hat{k}\cdot u(0)$ defined in Eq.\,\eqref{eq:firstintegral} as a constant of motion. The integral of motion for $\varepsilon \cdot u(\tau)$ given by Eq.\,\eqref{eq:secondintegral} yields
\begin{equation}
C_1(\tau) = - (\varepsilon \cdot u(\tau) - \varepsilon \cdot u(0)) = -\frac{e}{m}\mathcal{A}_0 \psi(\tau)\,.
\end{equation}
If we contract the ansatz Eq.\,\eqref{eq:ansatz} with the 4-vector $L^\mu$ we obtain for the coefficient $C_3(\tau)$
\begin{equation}
C_3(\tau) = h(\tau)\,.
\end{equation}
Finally, by invoking the condition $u^2 = u^2(0) = c^2$ we get for the coefficient $C_2(\tau)$
\begin{multline}
C_2(\tau) = \frac{1}{2}h^2(\tau)\hat{k}\cdot u(0)\\
+\frac{e}{m}\frac{\mathcal{A}_0 \psi(\tau)}{\hat{k} \cdot u(0)}\left(\varepsilon\cdot u(0) + \frac{1}{2}\frac{e}{m}\mathcal{A}_0 \psi(\tau) \right)\,.
\end{multline}
By substituting all the coefficients back to our ansatz Eq.\,\eqref{eq:ansatz} we obtain a final result
\begin{multline}\label{eq:usolution}
u^\mu(\tau) = u^\mu(0) -\frac{e}{m}\mathcal{A}_0 \psi(\tau)\varepsilon^\mu + \frac{1}{2}h^2(\tau) \hat{k} \cdot u(0)\hat{k}^\mu\\
+\frac{e}{m}\frac{\mathcal{A}_0 \psi(\tau)}{\hat{k} \cdot u(0)}\left(\varepsilon\cdot u(0) + \frac{1}{2}\frac{e}{m}\mathcal{A}_0 \psi(\tau) \right) \hat{k}^\mu\\
+ h(\tau) \epsilon^{\mu\nu\alpha\beta}u_\nu(0)\hat{k}_\alpha\varepsilon_\beta\;.
\end{multline}
It can be easily checked that the solution Eq.\,\eqref{eq:usolution} solves the dynamical equation given by Eq.\,\eqref{eq:firstplanewave}. Since a first order differential equation has only one solution, this is also a general solution for particle motion. Moreover, this solution has a very clear limit where if the magnetic moment charge $d_{m}$ vanishes, then $h(\tau)$ vanishes removing all effects of spin on the particle's motion. 

Once we set the magnetic dipole charge $d_m$ to zero we effectively uncouple the equations for particle motion and spin dynamics because Stern-Gerlach force is no longer present. In this case $h(\tau) \equiv 0$ and the solution Eq.\,\eqref{eq:usolution} reduces to
\begin{multline}\label{eq:volkovsolution}
	u^\mu(\tau) = u^\mu(0) - \frac{e}{m}\psi(\tau)\varepsilon^\mu\\
	+\frac{e}{m}\frac{\mathcal{A}_0 \psi(\tau)}{\hat{k} \cdot u(0)}\left(\varepsilon\cdot u(0) + \frac{1}{2}\frac{e}{m}\mathcal{A}_0 \psi(\tau) \right) \hat{k}^\mu\,,
\end{multline}
which is the well-known classical solution for a spinless charged particle in an external plane wave field~\cite{Itzykson2005}. 

We can easily evaluate the invariant acceleration as a square of Eq.\,\eqref{eq:firstplanewave}. After contraction of the antisymmetric tensors using Eq.\,\eqref{eq:contraction} we get
\begin{multline}\label{eq:invacc}
\dot{u}^2(\tau) = - \frac{\mathcal{A}_0^2\omega^2}{m^2c^2} (\hat{k}\cdot u(0))^2 \bigg(e^2f'(\xi)^2 \\ 
\left. + d_m^2\frac{\omega^2}{c^2}f''(\xi)^2 (\hat{k} \cdot s(\tau))^2 \right)\;. 
\end{multline}
This expression depends only on the solution for $\hat{k} \cdot s(\tau)$ defined by Eq.\,\ref{eq:kssolution} and is manifestly negative as $\dot{u}^\mu$ is a space-like vector. The invariant acceleration is therefore only a function of alignment and orientation of the spin 4-vector and the plane wave 4-vector which makes physical sense as the Stern-Gerlach force is sensitive to the alignment of the spin to the external magnetic field.

Finally, in order to assert uniqueness of solution (\ref{eq:usolution}), we would like to comment on a basis set which could be constructed in the Minkowski spacetime from the available 4-vectors. A good start would be a selection
\begin{align}
&u^\mu(0), \quad s^\mu(0)\,\nonumber\\
&F^\mu \equiv \epsilon^{\mu\nu\alpha\beta}u_\nu(0)\hat{k}_\alpha s_\beta(0)\,\nonumber\\
&G^\mu \equiv \epsilon^{\mu\nu\alpha\beta}u_\nu(0)\varepsilon_\alpha s_\beta(0)\,.
\end{align}
These four 4-vectors are all mutually orthogonal except for the product
\begin{equation}
F\cdot G = (\varepsilon \cdot s(0))(\hat{k}\cdot s(0))c^2 + (\varepsilon \cdot u(0))(\hat{k}\cdot u(0))s^2\,,
\end{equation}
which allows us to use Gramm-Schmidt orthogonalization to define a new 4-vector $H^\mu$ given by
\begin{equation}
H^\mu \equiv G^\mu - F\cdot G \frac{F^\mu}{F^2}\,,
\end{equation}
which together with $u^\mu(0)$, $s^\mu(0)$ and $F^\mu$ forms an orthogonal basis. This basis set becomes degenerate if either $F^\mu$ or $G^\mu$ is identically zero which happens if a quantity
\begin{equation}
\Omega \equiv \epsilon^{\mu\nu\alpha\beta}\hat{k}_\mu u_\nu(0)\varepsilon_\alpha s_\beta(0)\,,
\end{equation} 
is equal to zero. In that case another 4-vector has to be included to construct an orthogonal basis set dependent on the specific situation. In general we can always construct a basis composed of a time-like vector $u^\mu(0)$ and three other orthogonal space-like vectors.

Once the basis set is defined all 4-vectors can be expressed as linear combination of elements of such basis including the time-dependent 4-velocity $u^\mu(\tau)$. After solving the differential equations for the expansion coefficient we always recover the solution Eq.\,\eqref{eq:usolution} which we obtained by choosing the right ansatz. 

\subsection{Case of a particle initially at rest}\label{sec:restparticle}
\noindent In this section we will address the situation of a particle initially at rest $u^\mu(0) = (c,0,0,0)$ with respect to the laboratory observer. This case is of particular importance because it gives us an idea how the particle will react to the external field in the co-moving frame for situations involving a beam of charged particles subjected to a plane wave. For motion where the wave propagation and the particle beam are along the same axis, the results described here will differ only by the application of a Lorentz boost. Without loss of generality, we can choose to orient the coordinate system so that the wave unit vector is along the $z$-axis $\hat{\pmb{k}} = \hat{z}$ and the polarization unit vector is along the $x$-axis $\pmb{\varepsilon} = \hat{x}$. The initial spin is oriented in the arbitrary direction $\pmb{s}_0 = (s_{0x},s_{0y},s_{0z})$. The associated 4-vectors read
\begin{align}\label{eq:ourchoice}
\hat{k}^\mu = (1,&0,0,1), \quad \varepsilon^\mu = (0,1,0,0)\,,\nonumber\\
s^\mu(0) &= (0, s_{0x},s_{0y},s_{0z})\,.
\end{align}
\begin{figure}
	\includegraphics[width=0.9\linewidth]{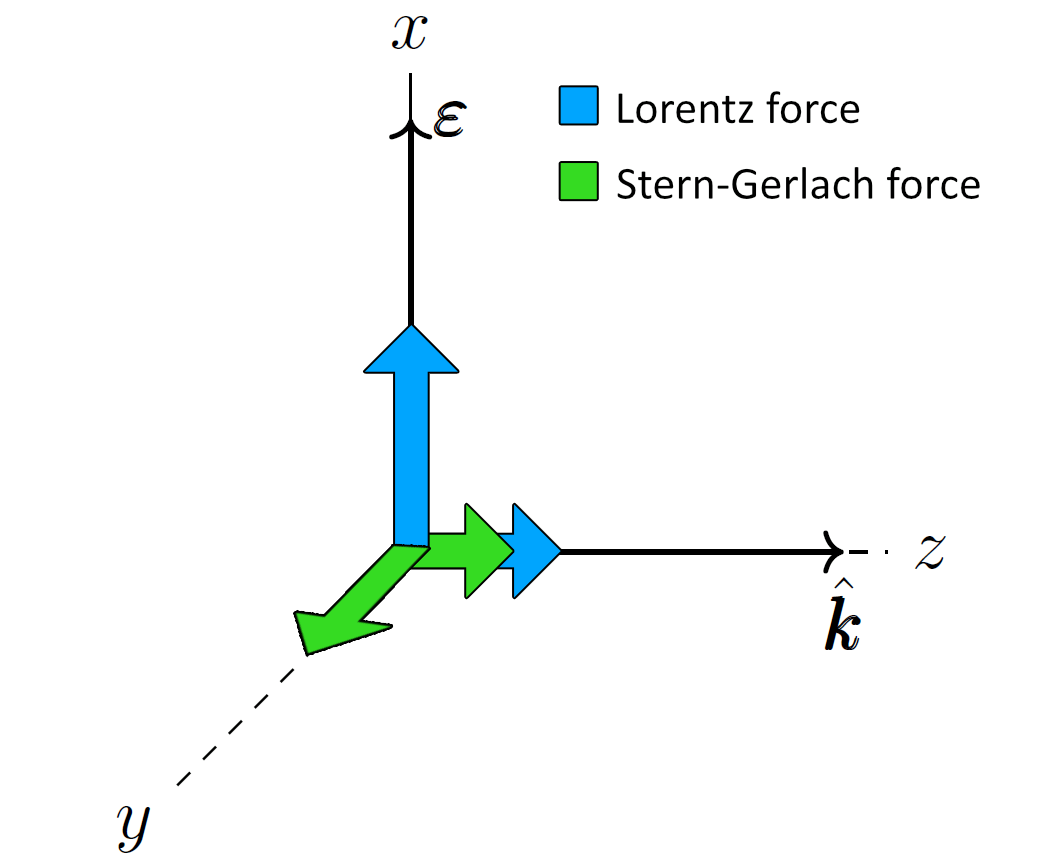}
	\caption{\label{fig:coordinates} Forces acting on an electron initially at rest by a plane wave with momentum $\hat{\pmb{k}}$ parallel to the $\hat z$-axis,  and polarized along the $\hat x$-axis. See Eq.~\eqref{eq:utau}. Both the Lorentz force (blue) and the Stern-Gerlach force  (green) induce motion along the plane wave in the $\hat{\pmb{k}}$ direction. Lorentz force causes additional oscillations along the polarization $\hat x$-direction. The Stern-Gerlach force causes a cumulative drift in the third orthogonal $\hat y$-direction -- sign depends  on the initial spin projection $s_{0x}$. Arrow lengths are  illustrative; however, for $a_0 \ll 1$ the transverse forces can dominate as shown.}
\end{figure}
Given the general 4-velocity solution from Eq.\,\eqref{eq:usolution}, we will get in the special case of particle initially at rest and plane wave described by 4-vectors in Eq.~\eqref{eq:ourchoice}
\begin{equation}\label{eq:utau}
u^\mu(\tau) = c \left(\begin{matrix}1 + \frac{1}{2}(h^2(\tau) + a_0^2 \psi^2(\tau))\\ -a_0 \psi(\tau)\\ h(\tau)\\ \frac{1}{2}(h^2(\tau) + a_0^2 \psi^2(\tau)) \end{matrix}\right)\,.
\end{equation}
In  here  the terms with $a_0\psi(\tau)$ are the standard solution for the Lorentz interaction with the plane wave fields, and the terms with $h(\tau)$ correspond to the magnetic moment interaction. We see that a particle initially at rest will move out into the direction normal to the plane wave propagation; we will speak of such velocity gain as  a drift velocity induced by the plane wave. For a better idea about geometry of this situation see Figure \ref{fig:coordinates}. We will devote the rest of this section to the study of the forces acting on the particle.  

We start by investigating the function $h(\tau)$ from Eq.~\eqref{eq:htau}, which governs the magnetic moment interaction. With our choice of the laser-particle configuration given by Eq.~\eqref{eq:ourchoice},  the function $W(0)$ from Eq.\,\eqref{eq:wdef} and projection $\hat{k}\cdot s(0)$ are
\begin{align}
W(0) &= -cs_{0x}\,,\\
\hat{k}\cdot s(0) &= - s_{0z}\,. 
\end{align}
These two constants control the spin projection $\hat{k}\cdot s(\tau)$ of Eq.\,\eqref{eq:kssolution} yielding
\begin{equation}
\hat{k}\cdot s(\tau) = - s_{0z}\cos \left[\frac{\widetilde{a}\mathcal{A}_0}{c}\psi(\tau)\right] + s_{0x} \sin \left[\frac{\widetilde{a}\mathcal{A}_0}{c}\psi(\tau)\right]\,.
\end{equation}
We see that this function is zero and remains zero in the case of the initial spin oriented only along the $y$-axis, i.e. for $s_{0x} = s_{0z} = 0$. When that happens the integral $h(\tau)$ Eq.\,\eqref{eq:htau} is identically zero and there is no Stern-Gerlach force acting on the particle. The solution for the particle's motion then reduces to the classical plane wave solution for the spinless electron seen in Eq.\,\eqref{eq:volkovsolution}.

The constants controlling the arguments of the sine and cosine functions can be rewritten for a charged particle as
\begin{equation}\label{eq:aa}
\frac{\widetilde{a}\mathcal{A}_0}{c} = \frac{ae\mathcal{A}_0}{mc} = - aa_0\,,
\end{equation}
where $a$ is the anomalous magnetic moment and $a_0 = |e|\mathcal{A}_0 / mc$ is the unitless normalized laser amplitude defined earlier in Eq.\,\eqref{eq:a0} as the parameter controlling the relativistic effects.

We will model the plane wave as a sine wave which is adiabatically switched on as the wave arrives at the particle position and then adiabatically switched off when the wave leaves. Throughout the motion the function $f(\xi)$ and all its derivatives are bounded by 1 and the initial condition at $\tau = 0$ is
\begin{equation}
f(\xi_0) = f'(\xi_0) = f''(\xi_0) = 0\,.
\end{equation}
With this, we have for $\psi(\tau)$ from Eq.\,(\ref{eq:psi})
\begin{equation}
\psi(\tau) = f(\xi(\tau)) - f(\xi_0) = f(\xi(\tau))\,.
\end{equation}
In this case, the integral for $h(\tau)$ Eq.\,\eqref{eq:htau} reads
\begin{multline}\label{eq:htausin}
h(\xi) = (1+a)a_0 \frac{\omega}{mc^2}\int_{\xi_{0}}^{\xi(\tau)} \bigg[-s_{0z}\cos[aa_0 f(\widetilde{\xi})]\\
- s_{0x}\sin[aa_0f(\widetilde{\xi})]\bigg] f''(\widetilde{\xi}) d\widetilde{\xi}\,,
\end{multline}
which we will split in the following analysis into two parts
\begin{equation}
h(\xi) \equiv h_1(\xi) + h_2(\xi)\,,
\end{equation}
corresponding to the first and second terms present in the integrand. For an electron, we typically have $aa_0 \ll 1$ in which case we can evaluate the first part of the integral in Eq.\,(\ref{eq:htausin}) as 
\begin{equation}\label{eq:h1tau}
h_1(\xi) = -(1+a)a_0 \frac{\omega s_{0z}}{mc^2}f'(\xi) + O(a^2a_0^2)\,,
\end{equation}
and we see that the function $h_1(\xi)$ is in this case oscillatory for the oscillatory wave. The absolute value of this expression can be bounded by 
\begin{equation}\label{eq:zestimate}
|h_1(\xi) (aa_0 \ll 1)| \leq (1+a) a_0 \frac{\omega |s_{0z}|}{mc^2} \approx a_0 \times 10^{-6}\,,
\end{equation}
where the value is given for the electron's initial spin align with the $z$-direction $s_{0z} = \pm \hbar/2$ and 1 eV laser light. Note that this term is present even for zero anomalous magnetic moment $a = 0$, but since it oscillates around zero it doesn't accumulate over many laser field oscillations.

The second part of the integral for $h(\xi)$ Eq.\,(\ref{eq:htausin}) can be evaluated in the lowest order in $aa_0$ as
\begin{equation}\label{eq:h2tau}
h_2(\xi) = -aa_0^2 \frac{\omega s_{0x}}{mc^2}\int_{\xi_0}^{\xi(\tau)} f(\widetilde{\xi})f''(\widetilde{\xi}) d\widetilde{\xi} + O(a^2a_0^3)\,.
\end{equation}
This integral starts accumulating only when the laser particle acquires a phase $\xi(\tau_a) = \xi_a$, $\tau_a$ being the time when the pulse arrives and the interaction is switched on. Neglecting the time interval of the laser plane wave ramp-on as short compared to duration of the pulse and approximating with $f(\xi) = \sin(\xi)$, we have
\begin{equation}
h_2(\xi) 
= a a_0^2 \frac{\omega s_{0x}}{mc^2}\left(\frac{\xi-\xi_a}{2} - \frac{1}{4}\sin2\xi\right) + O(a^2a_0^3)\,.
\end{equation}
The oscillatory part of this expression can be again bounded, this time by
\begin{equation}\label{eq:xestimate}
|h_2(\xi)(aa_0\ll1)|_\text{osc} \leq \frac{1}{4}aa_0^2 \frac{\omega |s_{0x}|}{mc^2}\approx \frac{1}{4}a a_0^2 \times 10^{-6}\,, 
\end{equation}
where the value is given for electron's spin along the $x$-direction $s_{0x} = \pm \hbar/2$ and 1 eV laser light. This contribution is much smaller than the $z$-direction spin polarization contribution Eq.\,(\ref{eq:zestimate}), because it linearly depends on the value of the anomalous magnetic moment $a$. Again, the oscillations are around zero and do not contribute over many plane wave periods. 

\begin{figure}
	\includegraphics[width=\linewidth]{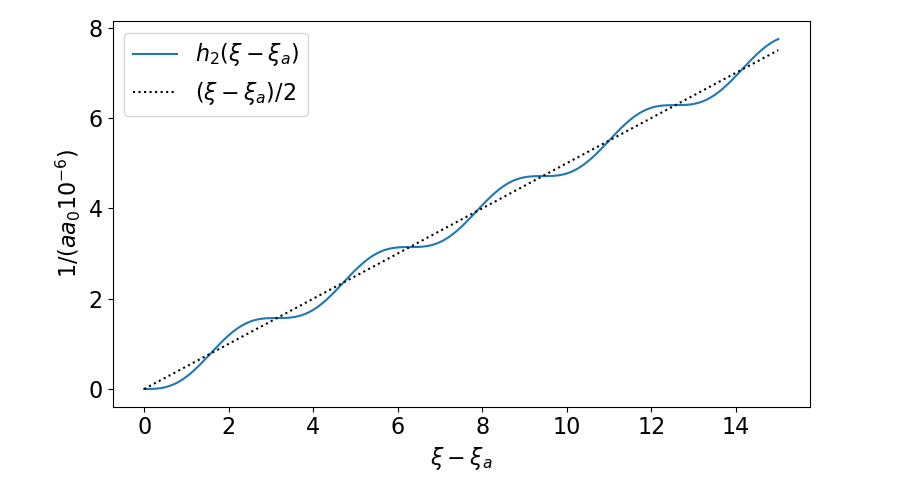}
	\caption{\label{fig:wiggle}The magnetically induced drift velocity $h_2(\xi)$ as given in the lowest order in $aa_0$ by Eq.\,(\ref{eq:h2tau}) for an electron spin in the $x$-direction. The dotted line depicts the overall trend in velocity increase.}
\end{figure}

The most important part of the function $h_2(\xi)$ is the linear term, which keeps accumulating over the interaction with many laser oscillations. The cumulative part is given by the expression
\begin{multline}\label{eq:h2taueq}
h_2(\xi)(aa_0 \ll 1)_\text{cum} = aa_0^2 \frac{\omega s_{0x}}{mc^2}\frac{\xi - \xi_a}{2} \\
= aa_0^2 \frac{\omega s_{0x}}{mc^2}\frac{\omega (\tau-\tau_a)}{2}\,,
\end{multline}
where the relationship between the phase and proper time Eq.\,(\ref{eq:phase}) was used. The plot of the whole function $h_2(\xi)$ from Eq.\,(\ref{eq:h2tau}) is presented in Figure~\ref{fig:wiggle}. We clearly see the overall linear trend. 

For an electron and 1 eV laser light we have
\begin{multline}\label{eq:h2tauacc}
h_2(\tau)(aa_0\ll 1)_\text{cum} \approx a a_0^2 \frac{\omega(\tau-\tau_a)}{2}\times 10^{-6}\\
 = \pi a a_0^2 N \times 10^{-6}\,,
\end{multline}
where 
\begin{equation}
N = \frac{\omega (\tau-\tau_a)}{2\pi}
\end{equation}
is the number of plane wave oscillations the particle interacted with before leaving the laser beam. This cumulative effect becomes dominant with respect to other contributions from Eqs.~(\ref{eq:zestimate},\ref{eq:xestimate}) when 
\begin{equation}
N > \frac{2}{\pi}\frac{a_0}{a}\,.
\end{equation}
For an electron with $a \approx 10^{-3}$ and with laser amplitude $a_0 = 0.1$, this happens after about 65 oscillations. When this condition is satisfied, the whole function $h(\tau)$ can be approximated just by the cumulative term (\ref{eq:h2tauacc}).

In the following, we will take $a_0 \ll 1$, which our $a_0 = 0.1$ roughly satisfies. In such a case, we can approximate the $\gamma$ factor from the zeroth component of Eq.~\eqref{eq:utau} with 
\begin{equation}
\gamma(\tau) = 1 + \frac{1}{2}(h^2(\tau) + a_0^2 \psi^2(\tau)) \approx 1\,,
\end{equation}
where we neglected the $h^2(\tau)$ and $a_0^2\psi^2(\tau)$ terms as negligible in the non-relativistic limit.

In the same limit we are going to neglect the drift motion in the $\hat{z}$-direction. The $a_0^2\psi^2(\tau)$ term corresponds to the intermittent particle acceleration/deceleration by the laser wave front in the direction of the wave vector. The $h^2(\tau)$ is a similar effect induced by the magnetic moment, this time cummulative with $a^2a_0^4N^2$ dependence. See Figure~\ref{fig:coordinates} for reference.

In the the direction of the polarization ($\hat{x}$), we have oscillatory 4-velocity caused by particle charge / plane wave interaction. This behavior has been already described in literature in detail~\cite{Rafelski2017,Esaray1993} and the velocity can be bounded by
\begin{equation}\label{eq:betax}
|\beta_x| \leq a_0\,.
\end{equation}
Although this velocity can be substantial, it does not accumulate, and it does not cause a drift in the trajectory, since the oscillations in velocity are around zero for a particle starting with zero velocity in $\hat{x}$-direction. 

Turning our attention to the magnetic moment contribution to 3-velocity, the drift velocity in the $\hat{y}$-direction can be approximated as
\begin{equation}\label{eq:betay}
\beta_y(\tau) \approx h(\tau) \approx a a_0^2 \frac{\omega \tau}{2} \times 10^{-6}\,,
\end{equation} 
where only the cummulative contribution Eq.~\eqref{eq:h2tauacc} is considered. 

The maximum velocity caused by the Lorentz force oscillations Eq.~\eqref{eq:betax} and the Stern-Gerlach drift velocity Eq.~\eqref{eq:betay} become comparable after
\begin{equation}
N \approx \frac{2}{\pi}\frac{10^6}{aa_0} \approx 10^{10}
\end{equation} 
oscillations. This would require keeping an  electron that was initially at rest  within the laser beam for about 20 $\mu$s, a challenging   laser beam control task.

Due to the cumulative Stern-Gerlach force, the $\hat{x}$-polarized electron drifts out from the typical laser beam radius $r_y = 1.5\, \mu\text{m}$ region after about
\begin{equation}
N \approx \sqrt{\frac{\omega r_y}{10^{-6}\pi^2caa_0^2}} \approx 3\times 10^5
\end{equation}
oscillations. During this time it acquires a transverse velocity in the $\hat{y}$-direction of approximately 3,000 m/s and the corresponding laser pulse length is roughly 1 ns.

The electron bunch is typically randomly polarized along the $\hat{x}$-direction, and the spin can have classically any value from $-\hbar/2$ to $\hbar/2$. The magnetic moment interaction described in this paper would result in a beam splitting along the $\hat{y}$-direction.

\section{Summary, Discussion and Conclusions}\label{Sec:Final}
\noindent 
In this work, we have added to the understanding of the contribution of the magnetic moment to electron dynamics in the presenece of an external EM plane wave field in an analytical fashion. Our classical model differs from the one used by Wen et al.~\cite{Wen:2017zer}, since we avoid introduction of particle mass modification. This model of particle motion when spin is involved was proposed a century ago by Frenkel~\cite{Frenkel1926} (for a reformulation in modern notation see~\cite{Ternov1980}), and disucssion of the spin dependent mass (even in homogeneous fields) is presented in~\cite{Kassandrov:2009jd}.

We have considered the two test cases of experimental relevance, which we can compare with the results of Wen et al.~\cite{Wen:2017zer}, who were using Frenkel mass modifying Stern-Gerlach force: (a) the motion of a particle traveling along the axis of a current loop in Section~\ref{sec:currentloop}, and (b) In Section~\ref{sec:planewave} the motion of a particle in EM plane waves.

(a) Motion in presence of current loop also in our approach leads to Stern-Gerlach trajectory splitting. We showed that electrons polarized in the direction of motion are delayed with respect to electrons with spin against the direction of motion. Our model qualitatively agrees with the classical limit of the DP equation. This result is also consistent with the KGP approach discussed above as the quantum KGP and DP equations are equivalent in the limit of \lq weak\rq\ external fields.

(b) We have explored motion of a charged particle in an external EM plane wave field. Previously, we presented an analytical solution for such behavior for neutral particles, where the magnetic moment interaction is a first order effect~\cite{Formanek:2019cga}. In this work we extend our solution to the case of charged particles and discuss implications for a particle initially at rest in the laboratory frame, see Section~\ref{sec:restparticle}. We focused on the case of a particle at rest, since this is the situation when a laser shot hits matter at rest. This case can be extended by means of a Lorentz transformation to incorporate another class of experimentally relevant situations of a particle beam moving parallel to the laser pulse. 

We showed that for a particle initially polarized in the direction of the plane wave's polarization, the Stern-Gerlach force pushes the particle in a direction perpendicular to both wave polarization and propagation. This would allow us to spatially separate electrons based on their polarization using a laser. 

Any electron beam consisting of particle bunches experiences intrinsic Coulomb repulsion forces in the transversal direction as well. This collective behavior beyond the dynamics of a single particle could overshadow the magnetic moment effect. However, this effect acts in a radial direction rather than along a plane. For a dedicated study of the spin Stern-Gerlach force, unbunched continuous beams are suggested in order to avoid the Coulomb driven beam spreading.

In this work we have not considered the process of emittion of radiation by electrons due to their motion in an external field. A possibility of the spin contribution to the electron radiation has been studied theoretically~\cite{Khokhonov2}, and demonstrated experimentally~\cite{kirsebom}. Here we draw attention to the expression for the invariant acceleration we obtained, see Eq. (\ref{eq:invacc}): the magnetic moment radiation expressed by acceleration squared can be  compared to electric dipole radiation. We see that magnetic acceleration strength acquires  an extra derivative of light wave $f^\prime\to f^{\prime\prime}$, and a cofactor $\omega/m$. This suggests (since expression is exact for plane wave and not for light pulses) that magnetic radiation can be comparable to electric dipole radiation strength considering particle within  highly singular light pulses.
	
The domain in which we have explored the Stern-Gerlach force is governed by classical physics
criteria as was  discussed in  Section \ref{sec:clasvsquant}. We argued in Ref. \cite{Rafelski:2017hce} that the \lq magnetic dipole\rq\ charge of a particle is a fundamental property alongside its rest mass and electric charge. We like to interpret the magnetic moment in terms of anomaly $a = (g - 2)/2$ since the effect that we describe depends on $a$. For an electron it so happens that the anomaly $a_e$, the deviation from Bohr magneton, is small and the magnitude is characterized by a fine structure constant and originates in the well-known Schwinger QED diagram. However, this should not be interpreted as if QED is part of the effects considered here.

That our results have no relation to quantum effects is best recognized by considering, instead of an electron, a proton, i.e., a particle with a magnetic moment that is quite different from (nuclear) Bohr magneton. In fact we do not expect any QED effects to appear in  particle dynamics in the  soft field of a continuous beam  laser, let alone to show cumulative effect we see for the Stern-Gerlach force spin dynamics.

However, it can be anticipated that more intense laser beams become available, and/or that we
port the physics we developed here to crystal channeling of electrons or/and protons. Therefore
in the future we would like to extend the magnetic moment interaction to the quantum domain by
incorporating the Stern-Gerlach potential into quantum-mechanical framework.  A useful tool on
this path would be a semi-classical treatment which shows a great promise to describe accurately
the ultra-relativistic motion~\cite{Bagrov, Wistisen}.

The dynamical examples presented demonstrate  that the electron beam control in some environments requires understanding and incorporation of the magnetic moment interaction due to the Stern-Gerlach type force in particle dynamics. Considering specific laser-particle initial configurations we have shown that the Stern-Gerlach force  due to plane (laser) wave  influences the velocity of charged particles in a cumulative way. This differs from the transverse effect due to the Lorentz force which primarily causes oscilatory motion. One can wonder if  this effect can be used to measure the anomalous magnetic moment of charged particles. Unlike the spin precession experiments, it would use the trajectory modification by Stern-Gerlach force, but a study of the achievable precision is still required. 

To conclude: in order to fully describe the behavior of electrons in external fields, the magnetic moment interaction cannot be neglected. We believe that our results will become relevant whenever electron beam control requires full account of the magnetic moment dynamics. 

\begin{acknowledgments}	
We would like to thank the anonymous referees for presenting several references helping our discussion of the parameters controlling the validity of the classical approach and addressing prior work.
\end{acknowledgments}


\end{document}